\documentclass[preprint,12pt,authoryear]{elsarticle}

\usepackage{amssymb}

\usepackage{graphicx}
\usepackage{multirow}
\usepackage{siunitx}
\usepackage{booktabs}

\usepackage[hyphens]{url}
\usepackage{hyperref}
\usepackage[hyphenbreaks]{breakurl}
\usepackage{breakurl}

\journal{arXiV}

\begin{document}

\begin{frontmatter}



\title{Automatic Cardiac Pathology Recognition in Echocardiography Images Using Higher Order Dynamic Mode Decomposition and a Vision Transformer for Small Datasets}


\author[inst1]{Andrés Bell-Navas\corref{cor1}}
\ead{a.bell@upm.es}

\author[inst1,inst2]{Nourelhouda Groun}
\ead{gr.nourelhouda@alumnos.upm.es}

\author[inst3,inst4]{María Villalba-Orero}
\ead{mvorero@ucm.es}

\author[inst3]{Enrique Lara-Pezzi}
\ead{elara@cnic.es}

\author[inst1,inst5]{Jesús Garicano-Mena}
\ead{jesus.garicano.mena@upm.es}

\author[inst1,inst5]{Soledad {Le Clainche}}
\ead{soledad.leclainche@upm.es}

\cortext[cor1]{Corresponding author}

\affiliation[inst1]{organization={ETSI Aeronáutica y del Espacio, Universidad Politécnica de Madrid}, 
            addressline={Pl. del Cardenal Cisneros, 3}, 
            city={Madrid},
            postcode={28040}, 
            country={Spain}}

\affiliation[inst2]{organization={ETSI Telecomunicación, Universidad Politécnica de Madrid},
            addressline={Av. Complutense, 30}, 
            city={Madrid},
            postcode={28040}, 
            country={Spain}}

\affiliation[inst3]{organization={Centro Nacional de Investigaciones Cardiovasculares (CNIC)},
            addressline={C. de Melchor Fernández Almagro, 3}, 
            city={Madrid},
            postcode={28029}, 
            country={Spain}}

\affiliation[inst4]{organization={Departamento de Medicina y Cirugía Animal, Facultad de Veterinaria - Universidad Complutense de Madrid},
            addressline={Av. Puerta de Hierro}, 
            city={Madrid},
            postcode={28040}, 
            country={Spain}}
            
\affiliation[inst5]{organization={Center for Computational Simulation (CCS)},
            city={Boadilla del Monte},
            postcode={28660}, 
            country={Spain}}

\begin{abstract}
Heart diseases are the main international cause of human defunction. According to the WHO, nearly 18 million people decease each year because of heart diseases. Also considering the increase of medical data, much pressure is put on the health industry to develop systems for early and accurate heart disease recognition. In this work, an automatic cardiac pathology recognition system based on a novel deep learning framework is proposed, which analyses in real-time echocardiography video sequences. The system works in two stages. The first one transforms the data included in a database of echocardiography sequences into a machine-learning-compatible collection of annotated images which can be used in the training stage of any kind of machine learning-based framework, and more specifically with deep learning. This includes the use of the Higher Order Dynamic Mode Decomposition (HODMD) algorithm, for the first time to the authors' knowledge, for both data augmentation and feature extraction in the medical field. The second stage is focused on building and training a Vision Transformer (ViT), barely explored in the related literature. The ViT is adapted for an effective training from scratch, even with small datasets. The designed neural network analyses images from an echocardiography sequence to predict the heart state. The results obtained show the superiority of the proposed system and the efficacy of the HODMD algorithm, even outperforming pretrained Convolutional Neural Networks (CNNs), which are so far the method of choice in the literature.
\end{abstract}


\end{frontmatter}


\section{Introduction}
\label{sec:intro}


Heart diseases or cardiovascular diseases (CVDs) are the main cause of human defunction in the world (\cite{Arooj2022deep}). According to the World Health Organization (WHO), a report conducted in 2019 shows that approximately 18 million people deceased that year precisely due to CVDs (\cite{who1999cvds}), supposing nearly a third of the total defunctions. Not only the number of patients is growing, but also medical data availability. Therefore, a lot of pressure is put on the health industry to develop systems for early and accurate detection of heart diseases, which usually takes much time, effort, and demanding resources. Researchers and medical experts mainly rely on images for the diagnosis, because of containing much information about the heart state useful for the further clinical procedures to adopt. Echocardiography imaging is very widely used, in particular transthoracic echocardiography (TTE), as being rapid and non-ionizing for diagnosis, therapeutic planning, and preoperative management of patients. The widespread use of mobile (handheld) echocardiography devices has significantly increased the recognition rate of heart diseases. However, several challenges concerning the quality and characteristics of the imagery (e.g., poor contrast) remain. These hamper the interpretations, which require the cardiologist's expertise.

On the other hand, as a tractable and repeatable modality, processing echocardiography images for diagnoses of cardiac pathologies through deep learning has become popular. This is because Artificial Intelligence (AI) has demonstrated to improve welfare and could make diagnoses automatic, quicker, more accurate and reproducible, reducing human mistakes and costs, and complementing decision-making processes. Several classification techniques have been proposed to help health-care experts in the diagnoses of heart diseases. For example, in \cite{Wahlang2021deep}, a Long Short Term Memory (LSTM) and a Variational Autoencoder (VAE) were proposed for binary classification (normal or abnormal) and classification between six different types of regurgitation in 2D echo, 3D Doppler, and videographic images. Similarly, in \cite{Madani2018deep}, a supervised model and a semi-supervised generative adversarial network (GAN) were used to classify left ventricle hypertrophy (LVH) and 15-view echocardiography images. The work in \cite{Vafaeezadeh2023carpnet} proposes CarpNet, a novel structure which aggregates spatio-temporal information by integrating CNNs and a Transformer, for the first time, for diagnosis of Mitral Valve (MV) leaflet abnormalities in four common Carpentier functional classes in parasternal long-axis (PLA) echocardiographic videos. Although the classification of these four classes using few test samples is challenging, Inception and ResNet version 2 with transfer learning for the CNN part of CarpNet have demonstrated to be superior to image-based algorithms using only spatial information. In \cite{farhad2023data}, novel zero-shot and few-shot learning based on a Siamese Neural Network were proposed for the first time for classification of LVH and healthy states using only 70 two-chamber (2CH) and four-chamber (4CH)-view echocardiography images. Contrarily to traditional zero-shot learning, this work does not require text vectors for image description. A shallow CNN was used as feature extractor, and the Euclidean distance was employed to compute the similarity between echo images. The proposed Siamese Network is superior to VGG16 and ResNet50 with transfer learning. The work from \cite{holste2022self} proposes self-supervised learning for effective representation learning for downstream fine-tuning on the task of automatic severe aortic stenosis (AS) diagnosis in echocardiogram videos with very little labeled data. To this end, positive pairs of different videos from the same PLAX view of the same patient are formed as a multi-instance contrastive learning. In this way, strong data augmentation which could potentially degrade signal is not required. A 3D ResNet18 is employed for the classification. In addition, the pretext task of frame re-ordering is used to incorporate temporal information, further aiding downstream fine-tuning, via a fully-connected classification head. The work in \cite{chen2023weakly} addresses hypertensive cardiomyopathy (HTCM) detection in echocardiography videos from 4CH view as a weakly supervised learning problem with a multi-instance deep learning framework. It uses a CNN for effective incorporation of temporal correlated multi-instance learning, end-systole and end-diastole attributes. In addition, the work employs domain adversarial neural networks for systematic addressing of variable appearance due to noise, blur, contrast, etc. In this way, the snippet most representative of the heart state in the video becomes greatly discriminative. Using a non-local I3D model pre-trained on Kinetics-400 data set as the 3D CNN feature extractor, the approach has demonstrated to outperform the vanilla I3D algorithm.  

Among the aforementioned works, CNNs are mostly proposed and usually focus on either binary classification (i.e., one specific disease), or heart state categories related to the same disease (like in \cite{Vafaeezadeh2023carpnet}). Besides, works proposing alternative algorithms, e.g., Transformers, either use very simple networks or vanilla versions of the algorithms, without any further contributions.

In this paper, a different solution is proposed, based on deep learning methods for recognition of different cardiac pathologies from echocardiography images. A novel procedure has been devised to create a large database from different sources of echocardiography images with heart diagnosis provided by human experts. These sources are heterogeneous regarding the image resolution and frame-rate, due to different acquisition conditions and sensors used. Therefore, it is not possible to use them simultaneously to create a larger database. The proposed procedure enables the homogenization of the different sources of echocardiography images to create a larger database required by the proposed deep learning method. This also includes the use of a data-driven method, the Higher Order Dynamic Mode Decomposition (HODMD) (\cite{le2017higher}), for the first time to the authors' knowledge. Its purpose is to generate images with less noise and more discriminative features than the original data associated to the different cardiac conditions, thus augmenting the training database and improving recognition accuracy. In fact, the HODMD modes contain temporal information, extracted from the sequences, which could be relevant to characterize different heart states. Regarding the proposed deep neural network, it incorporates modules to palliate the usual need for large databases, especially in Vision Transformers (ViT) (\cite{Lee2021vision}). As a result, the proposed system can automatically estimate the heart state from echocardiography video sequences without the intervention of human experts during its operation. Moreover, that estimation can be performed in real-time. This work largely extends the preliminary study carried out in \cite{Bell2023HODMD} by proposing the new database creation procedure, the deep learning architecture, and giving experimental results about the cardiac pathology recognition performance. The source code used in this work will be incorporated into the next version release of the ModelFLOWs-app (\cite{hetherington2023modelflows}), which can be found in \cite{modelflowsapp2023}.

The organization of the paper is as follows. Section \ref{sec:system} describes the proposed cardiac pathology recognition system. Section \ref{sec:database} introduces the database created to assess the proposed system. Section \ref{sec:results} summarizes the obtained results with the proposed system and makes a comparison with other algorithms. Finally, conclusions are drawn in Section \ref{sec:conclusions}.

\section{Cardiac pathology recognition system}
\label{sec:system}

The cardiac pathology recognition system here proposed is aimed to analyse echocardiography images using an adapted neural network architecture to predict the heart states in real-time. The neural network is based on a ViT that introduces modules to palliate the need for large training databases. Therefore, the ViT can be effectively trained from scratch, even on small datasets. The system output is the prediction of the heart state from the input sequence of echocardiography images. The cardiac pathology recognition system has been carefully trained using a new and large cardiac pathology database, whose creation is a key contribution of this paper. The medical database creation consists of a multi-stage procedure applied to sequences of echocardiography images showing hearts with different states and pathologies. The designed procedure not only generates a machine-learning compatible medical database, but also deals with the fact that the echocardiography images have been acquired with different sensors, and therefore they have different characteristics (image resolution and frame-rate, among others). On the other hand, it alleviates the need to collect large databases, as ViTs are data-hungry (\cite{Lee2021vision}).

\begin{figure}[!ht]

\centering
\includegraphics[width=8 cm]{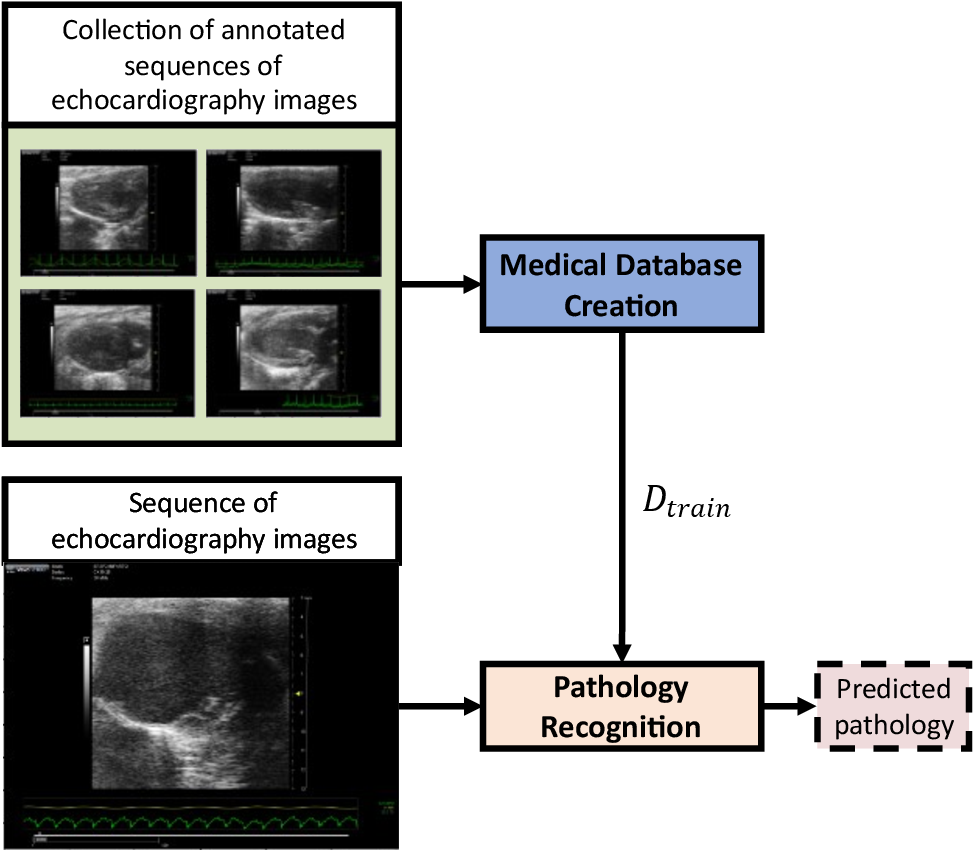}
\caption{Block diagram of the proposed cardiac pathology recognition system, with representations of the echocardiography images with non-heart regions (the electrocardiogram, medical information, and the black background).}
\label{fig:sys}

\end{figure}

Fig. \ref{fig:sys} depicts the block diagram of the proposed cardiac pathology recognition system, which is divided into two major stages: Medical Database Creation, and Pathology Recognition; those are described in detail in the next subsections.

\subsection{Medical Database Creation}
\label{sec:medical_data_creation}

\begin{figure}[!ht]
\centering
\includegraphics[width=14 cm]{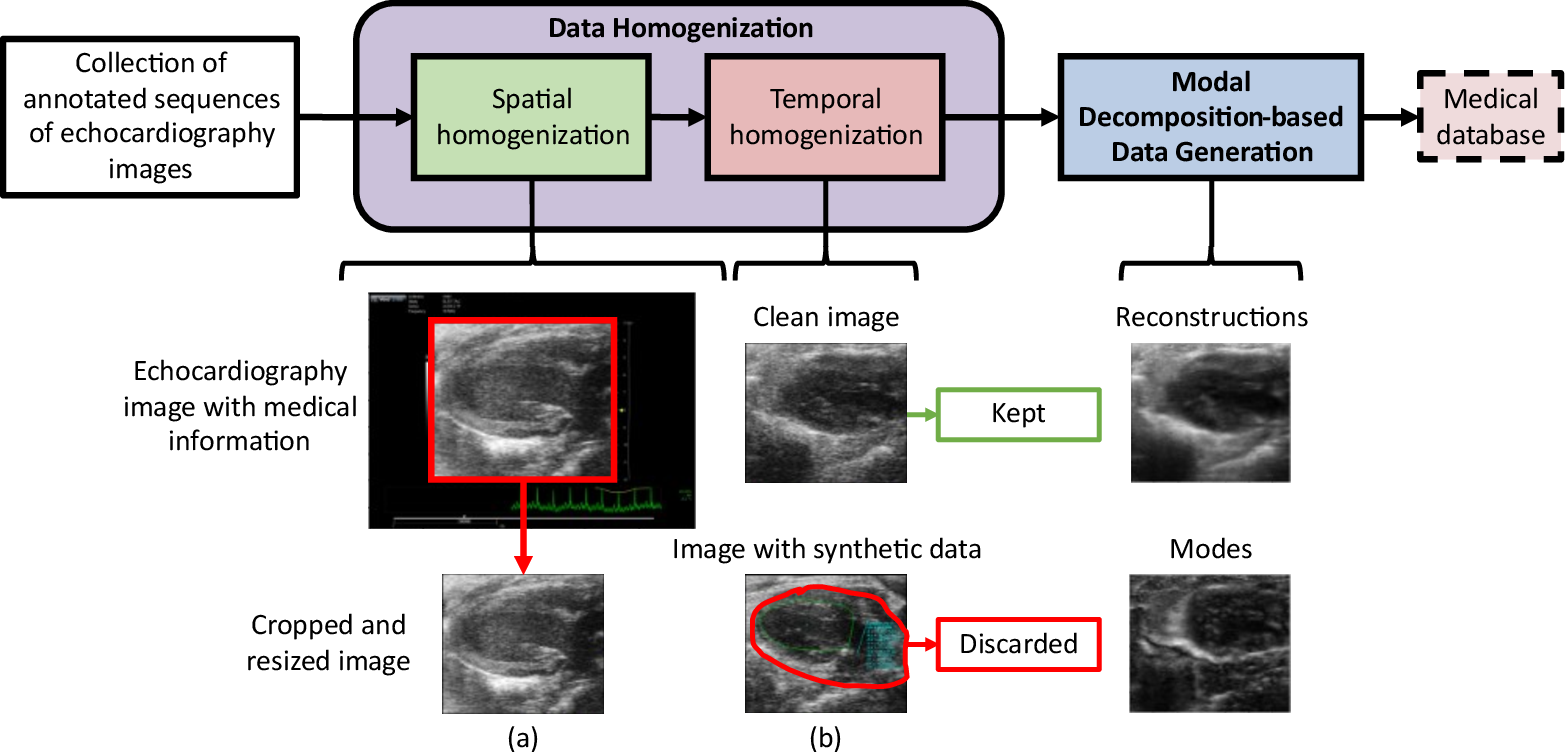}
\caption{Block diagram of the Medical Database Creation stage, composed of two phases: Data Homogenization, and Modal Decomposition-based Data Generation. The former phase is further divided into (a) Spatial homogenization and (b) Temporal homogenization.}
\label{fig:medical_data_creation}
\end{figure}

This stage creates a large annotated database from echocardiography images acquired with different sensors. The input consists of a heterogeneous collection of sequences of echocardiography images with additional medical information. Each sequence has the diagnosis of the heart state made by doctors. The output of this stage is a database composed of samples (as images) representing different heart states with homogenized spatial and temporal characteristics, adjusted to a target acquisition device. In this way, the resulting database can be used by machine learning algorithms. This Medical Database Creation stage can be further divided into two phases, as shown in Fig.~\ref{fig:medical_data_creation}: 1) Data Homogenization, and 2) Modal Decomposition-based Data Generation.

The first phase, Data Homogenization, extracts the images from the sequences, in the way that only the region representing the heart is retained, and all areas outside are discarded (containing medical information and the black background). This is achieved by using edge detection algorithms which locate the boundaries of the area corresponding to the echocardiography image (i.e., containing the heart). After that, the images containing text and graphics superposed on the heart area are filtered out. This is done to remove outliers, as having artificial data which do not represent the heart. Therefore, these should not be features associated to different heart states. For this purpose, Optical Character Recognition (OCR) has been adopted. This is because the artificial data in the studied databases always contain alphanumeric characters. Note that this makes sequences to have frames with different time intervals. However, in the second phase, Modal Decomposition-based Data Generation, the time interval between consecutive frames in a sequence must remain the same. This is accomplished by splitting a sequence with frames removed into sub-sequences, each one with the maximum possible number of consecutive frames with the same time interval. Each sub-sequence is therefore reinterpreted as an individual sequence of echocardiography images. 

In the second and last phase, Modal Decomposition-based Data Generation, novel samples are generated from the homogenized sequences. These consist of images with more discriminative features than those of the original echocardiography data. This generation is done by sequentially applying the Singular Value Decomposition (SVD) (\cite{sirovich1987turbulence}) and the HODMD algorithms (\cite{le2017higher}, \cite{Groun2022higher}) on each homogenized sequence, and taking from the outputs of each algorithm. Therefore, this phase can be seen as a machine learning method based on the data physics, in which data augmentation with modal decomposition is performed. 

The steps performed in the Modal Decomposition-based Data Generation phase are as follows. First, the SVD algorithm is employed. This generates a series of modes and the reconstructions of the echocardiography images. Next, the HODMD algorithm is applied on these reconstructions obtained with the previous SVD algorithm. As a result, HODMD modes and reconstructions associated to the sequence are obtained. Subsubsection \ref{sec:hodmd} describes in more detail the HODMD algorithm for feature extraction and data augmentation. Finally, the SVD algorithm is applied on the HODMD modes, obtaining SVD modes associated to the HODMD modes, and their reconstructions. From the data obtained in this latest use of the SVD algorithm, only the reconstructions of the HODMD modes are considered, and the modes are discarded. Note that a sequence must cover a number of heart cycles enough to be able to apply the HODMD algorithm and, therefore, the subsequent SVD algorithm. In this way, the temporal information is correctly characterized, and fair reconstructions can be obtained. This is accomplished by applying a threshold which indicates the minimum number of snapshots required in a sequence to apply the HODMD algorithm. Note that the reconstructions are leveraged due to having less noise than their associated original data, making the patterns of the different heart pathologies more distinct. In addition, the deep neural network can potentially learn better the temporal information of the sequences with the use of the obtained HODMD modes and their associated reconstructions (obtained with the second use of the SVD algorithm). It is due to that the HODMD modes describe the physical patterns associated to the dynamics of the data. These physical patterns can be related to heart diseases (\cite{Groun2022higher}).

As a result, an annotated database with a large number of cardiac pathology samples is obtained, whose spatial and temporal characteristics are adjusted to the target sensor used in the proposed recognition system. This is used as a training database $D_{train}$ to learn a deep neural model, which is the core of the Pathology Recognition stage described in Subsection \ref{sec:pat_recog}. Table~\ref{tab:training_cases} summarizes the different combinations of the data generated by the SVD and HODMD algorithms which are taken to form $D_{train}$. Precisely, the combinations with HODMD modes and reconstructions of the echocardiography images lead to the best results, because of having more samples and also more discriminative features derived from modal decomposition.  

\subsubsection{Higher Order Dynamic Mode Decomposition}
\label{sec:hodmd}

The Higher Order Dynamic Mode Decomposition algorithm (HODMD) (\cite{le2017higher}) is an extension of the Dynamic Mode Decomposition (DMD) (\cite{schmid2010dynamic}), extensively used in fluid dynamics and for diverse industrial applications (\cite{Groun2022higher}, \cite{VEGA202129}). HODMD decomposes spatio-temporal data (in this case, a video sequence of echocardiography images) into a number of DMD modes. Each one is associated to a frequency, growth rate, and amplitude. These modes represent the most discriminative patterns associated to the different heart states in echocardiography images. This also allows to identify and filter out noisy modes. In this way, reconstructions of each frame from the input sequence are obtained, with less noise and with more discriminative features associated to the different heart states. The HODMD algorithm used in this work is specifically the multidimensional iterative HODMD algorithm (\cite{LECLAINCHE2017336}, \cite{Groun2022higher}). It is based on the Higher Order Singular Value Decomposition (HOSVD) (\cite{tucker1966some}), which extends the SVD algorithm by applying it along each spatial dimension to better clean the data.

\begin{figure*}[!ht]
\centering
\includegraphics[width=14 cm]{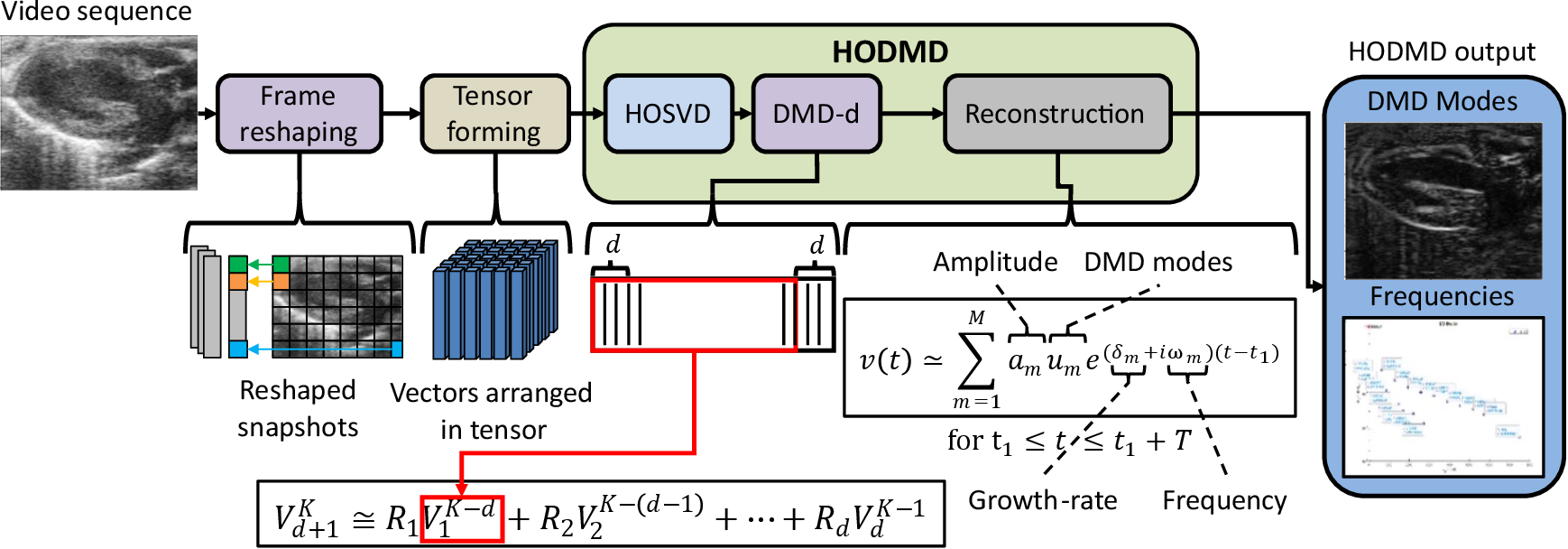}
\caption{Block diagram of the HODMD algorithm applied on a sequence of echocardiography images.}
\label{fig:hodmd}
\end{figure*}

Fig.~\ref{fig:hodmd} depicts the steps performed by the HODMD algorithm on a sequence of echocardiography images, which are as follows. Assuming that the sequence has been homogenized as described in Subsection \ref{sec:medical_data_creation}, each frame (or snapshot) is reshaped into a vector of dimensions $N_{p} = N_{x} \times N_{y}$, where $N_{x}$ and $N_{y}$ denote the spatial resolution of the snapshots, and $N_{p}$ the number of pixels of the snapshot. Once all vectors have been computed, a tensor is formed. The dimensions of the tensor are $N_{p} \times K$, where $K$ is the number of snapshots of the sequence. Then, a dimensionality reduction is carried out by applying the HOSVD algorithm. After that, a decomposition in eigenvalues and eigenvectors is conducted. In this case, a number of $d$ subsequent snapshots is considered as a contribution given by the HODMD algorithm to improve the precision. Next, the DMD modes are computed. Finally, a mode expansion process is performed, obtaining the frequencies, the growth-rate, and the amplitudes. All these steps are applied iteratively until the number of HOSVD modes converges, i.e., remains the same. The DMD modes allow to eventually reconstruct each snapshot (each frame) from the input sequence. The most representative DMD modes, containing characteristic patterns associated to the different heart states, together with the reconstructions, can be leveraged by the deep neural network to improve the recognition performance.

\subsection{Pathology Recognition}
\label{sec:pat_recog}

The Pathology Recognition stage predicts heart states from the corresponding input echocardiography images in real-time. Specifically, the input is a sequence of echocardiography images obtained by medical devices. It is adapted to the target acquisition device used in production in the previous Medical Creation Stage. The output of the Pathology Recognition stage is a cardiac pathology prediction which indicates the most probable heart state.

\begin{figure*}[!ht]
\centering
\includegraphics[width=14 cm]{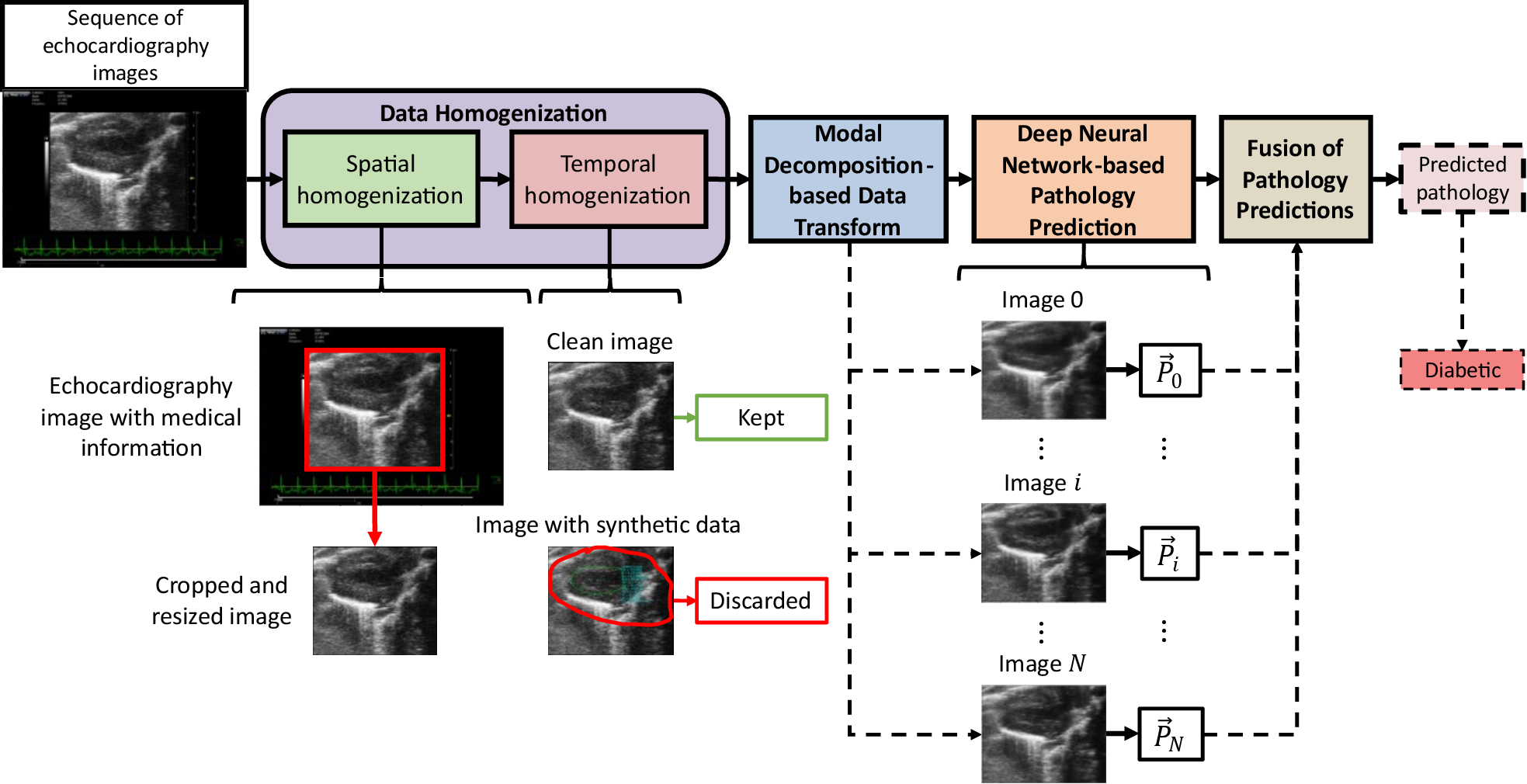}
\caption{Block diagram of the Pathology Recognition stage.}
\label{fig:pred}
\end{figure*}

The Pathology Recognition stage can be divided into four phases, as can be seen in Fig. \ref{fig:pred}: 1) Data Homogenization, 2) Modal Decomposition-based Data Transform, 3) Deep Neural Network-based Pathology Prediction, and 4) Fusion of Pathology Predictions. In the first phase, Data Homogenization, the input sequence is homogenized as already described in Subsection \ref{sec:medical_data_creation}.

In the second phase, Modal Decomposition-based Data Transform, the same process as in the Modal Decomposition-based Data Generation phase described in Subsection \ref{sec:medical_data_creation} is performed. This is, images with more discriminative features than the original echocardiography data are obtained by applying the SVD or HODMD algorithms. Note that, in this case, the resulting images from this phase are exclusively used as input of the deep neural network. Therefore, this phase acts only as a feature extractor. However, the Modal Decomposition-based Data Generation phase leverages the output of each applied algorithm to enlarge the training database $D_{train}$.

The third phase, Deep Neural Network-based Pathology Prediction, computes the cardiac pathology prediction for each image generated in the previous phase. For this purpose, a deep neural network architecture based on the ViT is used. It incorporates the Shifted Patch Tokenization (SPT) and Locality Self Attention (LSA) modules (\cite{Lee2021vision}) to effectively learn from medical data. These two modules jointly deal with the problem of lack of locality inductive bias inherent to ViTs, reducing the need of very large training datasets and improving performance.

\begin{figure}[!ht]
\centering
\includegraphics[width=12 cm]{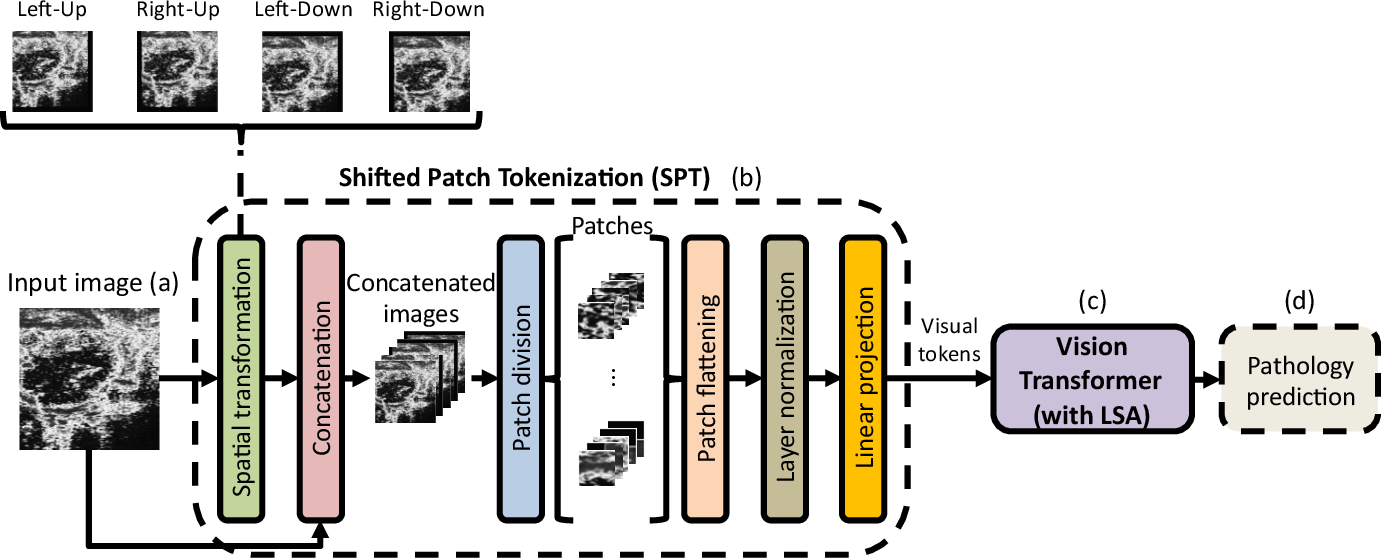}
\caption{Architecture of the proposed deep neural network. (a) The input: an image from an echocardiography video sequence. (b) The SPT module for more embedding of spatial information in each visual token. (c) The ViT with multi-head attention based on LSA for more locality inductive bias. (d) The predicted heart state of the input image.}
\label{fig:model}
\end{figure}

The architecture is represented in Fig. \ref{fig:model}. It is composed of the SPT module to exploit the spatial relations between neighboring pixels in the tokenization process, and a ViT adapted with LSA. The input is an image representing a heart in a determined state, and the output is a vector of score predictions, i.e., the confidences which represent each heart state. The processing of the input starts with a Shifted Patch Tokenization module, in which four diagonally shifted versions are first generated. Then, these images are concatenated with the input, and a division in non-overlapping patches is performed. Next, the spatial dimension of the patches is flattened, followed by a normalization layer and a linear projection. Before processing the patches with the ViT, a patch encoding layer is used to add positional information to them. By embedding more spatial information in each visual token, the locality inductive bias is increased, so the receptive field to visual tokens, improving standard tokenization, and thus the final performance of the ViT.

The backbone which processes the visual tokens is based on a ViT, whose multi-head attention incorporates LSA to increase the locality inductive bias. This has been done with eight stacked Transformer blocks, each one with the adapted multi-head attention layer, with four heads, projection dimension $64$ and dropout factor $10\,\% $. LSA is aimed to increase the locality inductive bias, making the attention of the ViT locally focused. This is done by sharpening the distribution of attention scores in two ways: one, by learning the temperature parameters of the Softmax function. And two, by removing the self-token relation with the use of diagonal masking. This forces to suppress the diagonal components of the similarity matrix computed by the query and key, increasing the attention scores between different tokens.  

The adapted multi-head attention layer is followed by an empty branch (skipped connection) and two Transformer units, each one composed of a fully connected layer with the Gaussian Error Linear Unit (GELU) activation function. The use of skip connections addresses the vanishing gradient problem, inherent to deep neural networks, to considerably improve training convergence. The dropout factor is here also $10\,\% $, and the number of neurons of the dense layers has been set to $128$ and $64$, respectively. The Transformer block finally introduces a skip connection, connecting the output of the previous one to the output of the Transformer unit. After all Transformer blocks, features are flattened and introduced into a Dropout layer and two head units, each one composed of a fully connected layer with the GELU activation function. The dropout factor of these layers is $50\,\% $, and the number of elements of the dense layers is $1024$ and $512$, respectively. Finally, a fully connected layer is used, whose output is a vector of score predictions containing the confidences of representing each heart state.

The deep neural network uses the cardiac pathology database of medical images, $D_{train}$, obtained in the Medical Database Creation stage described in Subsection \ref{sec:medical_data_creation}. Concerning the optimization technique, the AdamW algorithm has been used. The employed learning rate strategy is based on the warm-up cosine policy (\cite{Lee2021vision}) that computes the learning rate at iteration $i$ by the expression $\lambda_{i} = 0.5 \times \lambda_{t} \times (1 + cos(\pi \times (i - N_{w}) / (N_{iter} - N_{w})))$. The target learning rate is fixed to $\lambda_{t}$ = 0.001, and the warmup steps $N_{w}$ to $10\% \times N_{iter}$, where $N_{iter}$ is the maximum number of iterations. The momentum and weight decay are set to 0.9 and 0.0001, respectively. The batch size has been set to 64 images due to the limitations of the physical memory of the used GPUs (Nvidia A100). About the database division, a splitting scheme of $60 \% - 20 \%$ for training and validation, respectively, has been adopted, balancing the cardiac categories to ensure that every set contains approximately the same number of samples for each heart state. The batch generation scheme minimizes the adverse effects of data imbalance by forcing a minimum proportion of samples of every class. In this way, the skewed predictions towards the heart state classes with more samples are prevented. The batch generation also includes Data Augmentation techniques based on resizing, random horizontal flips, random rotations and random zooms, applied to the echocardiography data. In this way, not only potential overfitting problems could be palliated, but also the model could become more robust to different perspectives of the scene.

\begin{figure}[!ht]

\centering
\includegraphics[width=8.5 cm]{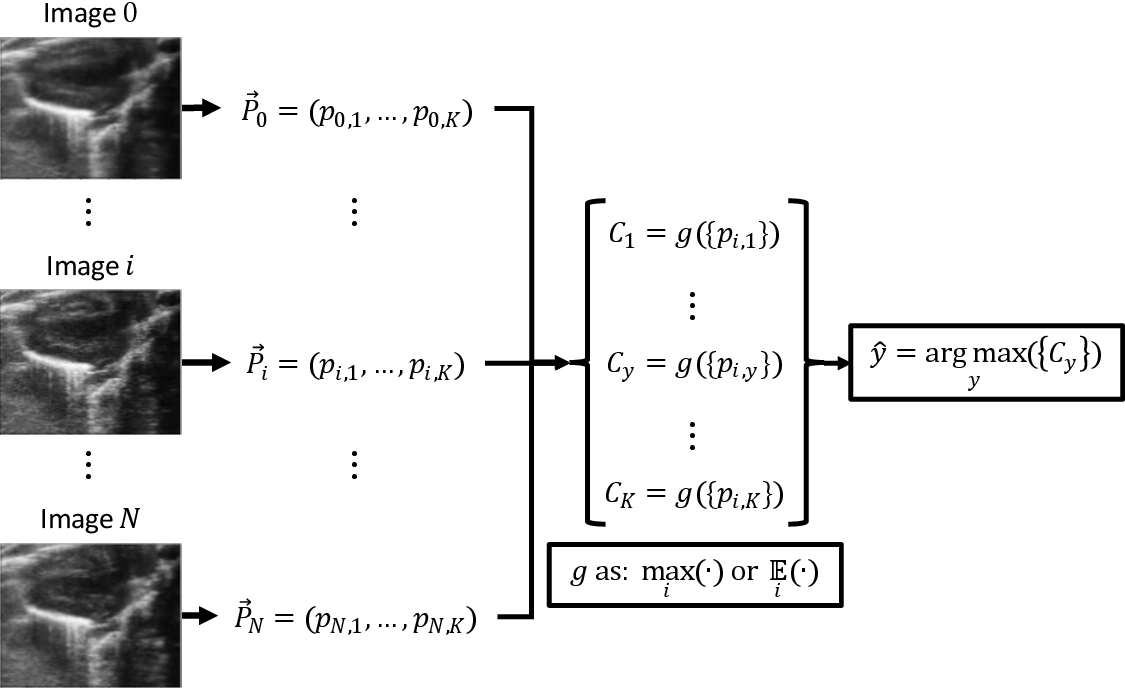}
\caption{Illustration of the Fusion of Pathology Predictions phase. The predictions corresponding to each heart state class from every image are fused, and the highest value determines the predicted heart state class of a test sequence.}
\label{fig:fusion}

\end{figure}

In the fourth and last phase, Fusion of Pathology Predictions, the most probable heart state of the input test sequence is determined by combining multiple predictions. This process is illustrated in Fig. \ref{fig:fusion}. Each image from the sequence has associated a vector of score predictions $\vec{P}_{i} = (p_{i,1}, ..., p_{i,K})$, that is, the confidences of belonging to each class. The set of scores $\{p_{i,y}\}$ is then used to compute a unique prediction score of the heart state $y$. For this purpose, the average and the maximum scores have been tested. Finally, the most probable heart state class is selected by $\hat{y} = \arg \max\limits_{y} (\{C_{y}\})$, as long as $\max \limits_{y} (\{C_{y}\})$ reaches a minimum threshold. Otherwise, the test sequence is classified as undetermined. Note that relying on a sequence of echocardiography images instead of a single one to predict pathologies is more robust, as it can help doctors to make more confident heart diagnoses. In addition, certain heart pathologies might lead to arrhythmias or patterns in the heart cycles which, therefore, constitute temporal information.

\section{Database}
\label{sec:database}

\begin{figure*}[!ht]

\centering
\includegraphics[width=8.5 cm]{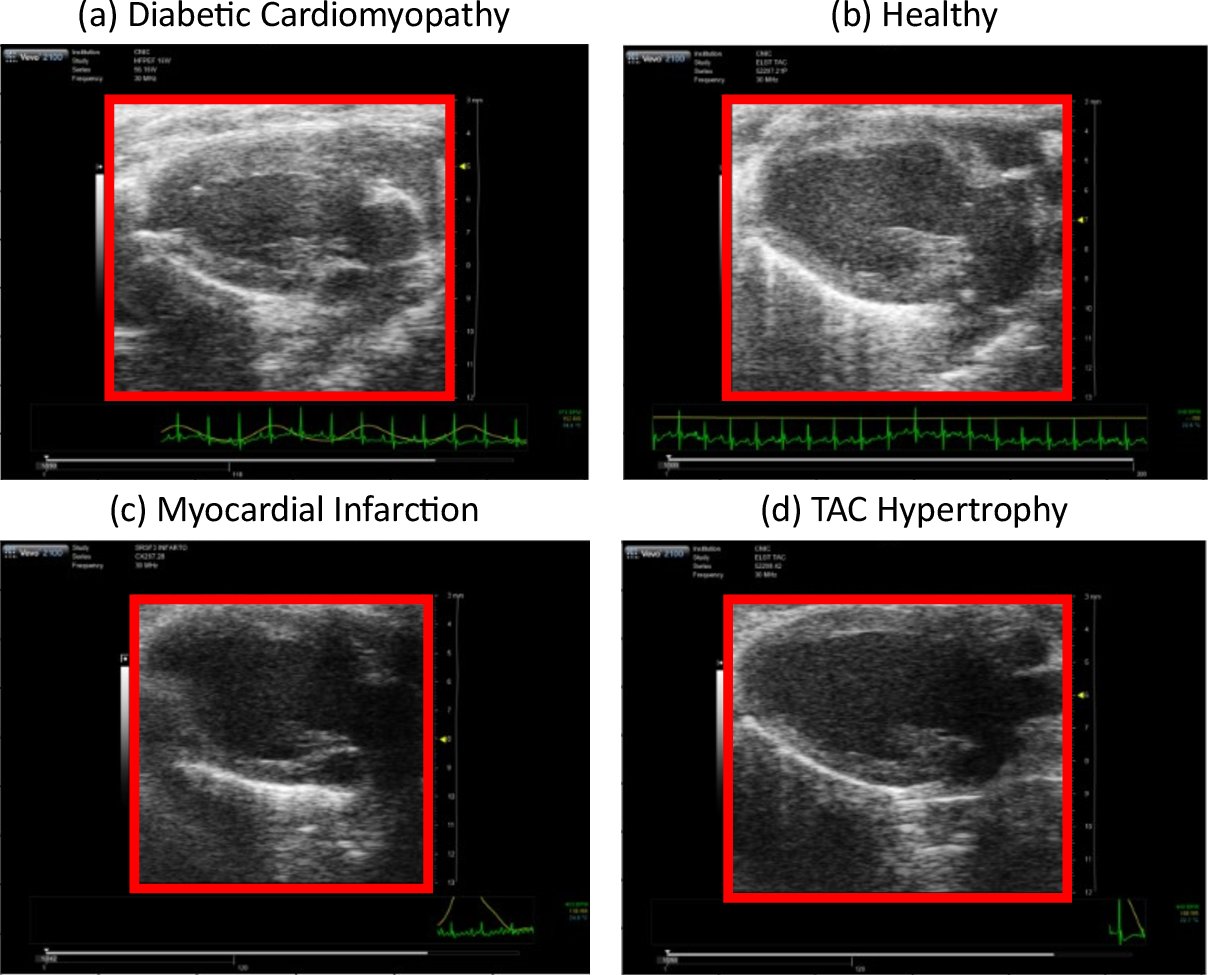}
\caption{Sample images from the different heart states of the echocardiography data. The regions of interest are delimited by the red boundaries.}
\label{fig:database}

\end{figure*}

The cardiac pathology recognition system here proposed has been tested and validated using a database of echocardiography data. It is composed of video sequences taken from several mice in diverse heart conditions based on mouse models, whose diagnoses were provided by experts: diabetic cardiomyopathy, healthy, myocardial infarction, and cardiac hypertrophy (TAC hypertrophy from now on). These models follow the protocols approved by the Centro Nacional de Investigaciones Cardiovasculares (CNIC) Institutional Animal Care and Research Advisory Committee of the Ethics Committee of the Regional Government of Madrid (PROEX177/17) (\cite{Groun2022higher}). Also, the performed experiments comply with the ARRIVE guidelines. Standard bidimensional (2D) PLAX images of the left ventricle of the heart were obtained and used in this work. The acquisition was performed based on TTE using a high-frequency ultrasound system (Vevo 2100, Visualsonics Inc, Canada) with a 40-MHz linear probe (\cite{Groun2022higher}). A sample image from each heart state is shown in Fig.~\ref{fig:database}. As can be seen, the echocardiography imagery has much noise, as expected in this modality of data. In addition, different perspectives of the LAX view are taken, boosting the complexity of the cardiac pathology recognition task. Only the area of the heart, i.e., the region of interest (ROI), is taken for the recognition system. The obtained heart areas have different resolutions, showing the heterogeneity of the echocardiography imagery: in mean and standard deviation, the spatial resolution is $659.10 \pm 54.80 \times 581.04 \pm 18.18$ pixels. However, the aspect factor among the images barely changes and it is close to 1, so these are practically square: $1.13 \pm 0.09$. Therefore, resizing to square images for the deep neural networks barely deforms the heart areas.

A summary of the main characteristics of the cardiac pathology database is presented in Table~\ref{tab:lax_data}. It is composed of the original echocardiography images, and the associated SVD-based data and the HODMD-based data. The latter were obtained after applying the HODMD algorithm on the SVD reconstructions of the original images. Note that the total number of original snapshots is also the number of associated reconstructed ones after applying the SVD algorithm, but the amount of reconstructions obtained after using the HODMD algorithm is less or equal. This is due to the minimum number of frames required in a sequence in order to apply the HODMD algorithm, which is not achieved in some of the sequences. Note that the training, validation and test sets comprise different sequences, which are the ones that both achieve the splitting scheme proposed in this work and a reasonable balance in the number of samples in each class.

\begin{table*}[!ht]
\centering
\caption{Summary of the main characteristics of the cardiac pathology database.}
\label{tab:lax_data}
\resizebox{\textwidth}{!}{%
\begin{tabular}{@{}ccccccc@{}}
\toprule
                                         &            &              &              &              & \multicolumn{2}{c}{HODMD-based Data} \\ \midrule
Heart State                              & Set        & $\#$ Sequences & $\#$ Snapshots & $\#$ SVD Modes & $\#$ Modes      & $\#$ Reconstructed     \\
\midrule
\multirow{3}{*}{Diabetic Cardiomyopathy} & Training   & 20           & 2276         & 100          & 1137          & 2214                 \\
                                         & Validation & 9            & 796          & 45           & 408           & 735                  \\
                                         & Test       & 10           & 785          & 50           & 351           & 715                  \\
\multirow{3}{*}{Healthy}                 & Training   & 21           & 2382         & 105          & 1176          & 2341                 \\
                                         & Validation & 10           & 893          & 50           & 480           & 832                  \\
                                         & Test       & 8            & 871          & 40           & 405           & 837                  \\
\multirow{3}{*}{Myocardial Infarction}   & Training   & 19           & 2425         & 95           & 1092          & 2425                 \\
                                         & Validation & 6            & 900          & 30           & 360           & 900                  \\
                                         & Test       & 6            & 900          & 30           & 345           & 900                  \\
\multirow{3}{*}{TAC Hypertrophy}        & Training   & 19           & 2243         & 95           & 960           & 2443                 \\
                                         & Validation & 6            & 720          & 30           & 333           & 720                  \\
                                         & Test       & 6            & 720          & 30           & 345           & 720                  \\ 
\multicolumn{2}{c}{Total}                             & 140          & 15911        & 700          & 7392          & 15582                \\ \bottomrule
\end{tabular}%
}
\end{table*}

\begin{table*}[!ht]
\centering
\caption{Summary of the combinations for $D_{train}$ used in the Medical Database Creation stage.}
\label{tab:training_cases}
\resizebox{\textwidth}{!}{%
\begin{tabular}{@{}ccccccc@{}}
\toprule
     &          & \multicolumn{2}{c}{SVD-based Data} & \multicolumn{2}{c}{HODMD-based Data} & SVD (HODMD Modes) \\ \midrule
Case & Original & Reconstructed       & Modes        & Reconstructed        & Modes         & Reconstructed   \\
\midrule
1    & $\times$   &                     &              &                      &               &                 \\
2    &          & $\times$              &              &                      &               &                 \\
3    & $\times$   & $\times$              & $\times$       &                      &               &                 \\
4    & $\times$   & $\times$              &              &                      &               &                 \\
5    &          & $\times$              & $\times$       &                      &               &                 \\
6    &          &                     &              & $\times$               &               &                 \\
7    &          &                     &              & $\times$               & $\times$        &                 \\
8    &          & $\times$              &              & $\times$               &               &                 \\
9    & $\times$   & $\times$              &              & $\times$               &               &                 \\
10   &          &                     &              & $\times$               & $\times$        & $\times$          \\
11   &          & $\times$              & $\times$       & $\times$               & $\times$        &                 \\
12   &          & $\times$              & $\times$       & $\times$               & $\times$        & $\times$          \\ \bottomrule
\end{tabular}%
}
\end{table*}

Table~\ref{tab:training_cases} shows the experimented cases of the training database $D_{train}$, each one indicating which types of data are used to form that. In this way, the impact of each type of data on the final performance could be evaluated. The final column represents the use of the reconstructions of the HODMD modes after applying the SVD algorithm on them.

\section{Results}
\label{sec:results}

The algorithm here considered has been tested with the cardiac pathology database using the different cases for the training database $D_{train}$. Additionally, it has been compared with CNN-based algorithms, including Inception-version 3 (\cite{szegedy2016rethinking}), ResNet50-version 2 (\cite{he2016identity}), and VGG16 (\cite{simonyan2014very}). The following metrics have been utilized to measure the performance of each algorithm in terms of classification accuracy and computational cost: F1-Score (\textit{F}), accuracy, and average processing time per image $\bar{t}$. The F1-Score is the harmonic mean of the precision and the recall. This is, the ratio of correct detections, that is, true positives (TP): samples correctly classified as a determined heart state, with respect to the correct detections and the mean of the false positives (FP): samples wrongly detected as the heart state, and false negatives (FN): actual samples from the heart state wrongly classified. The F1-Score is given by:  

\begin{equation}
\centering
F = \frac{\mathrm{TP}}{\mathrm{TP} + \frac{1}{2}(\mathrm{FP} + \mathrm{FN}))}
\end{equation}

The accuracy is the percentage of the correctly predicted samples (either images or videos) with respect to the total number of samples:

\begin{equation}
\centering
Accuracy = \frac{\#\:Correct\:predictions}{\#\:Samples}    
\end{equation}

To represent global values of accuracy, three types have been used: the per-image accuracy (w/o), the per-image accuracy after the Fusion of Pathology Predictions
phase (w/), described in Subsection \ref{sec:pat_recog}, and the per-sequence accuracy. The per-image accuracy is the one which considers each sample as an image. The one after the Fusion of Pathology Predictions phase measures the accuracy when the samples (as images) are classified as the predicted heart state in its associated sequence. Finally, the per-sequence accuracy considers each sample as a sequence. 

The average processing time per sample $\bar{t}$ is the average time taken by the proposed system to process a sample (either image or sequence) and give the heart state classification. It is broken down into the SVD algorithm ($\bar{t}_{SVD}$), the HODMD algorithm, comprising the HOSVD dimensionality reduction ($\bar{t}_{HOSVD}$) prior to the HODMD itself ($\bar{t}_{HODMD}$) (\cite{Bell2023HODMD}), and the Deep Neural Network-based Pathology Prediction phase ($\bar{t}_{pred}$). Unless the contrary is explicitly specified, the times are measured in milliseconds.  

The results have been obtained using a cluster with Intel Xeon Gold 6240R and three Tesla A100 GPU working in parallel for training and Intel Xeon Gold 6230 and one Tesla V100 GPU for testing. In this way, not only the high amount of data conforming the created medical database is managed, but also the training convergence is sped up and eased.

\subsection{Analysis of types of data}
\label{sec:analysis_data}

First, a study to evaluate the performance in terms of accuracy of the training cases on different types of data used for test is presented. To that end, the configuration of the SVD and HODMD algorithms, the deep neural network, and of the prediction is initially introduced. Later on, the influence of the types of data used for both training and test on the overall performance of the system is studied. Lastly, the performance in each heart state for the best configuration is shown.

\begin{enumerate}
\item Configuration: Regarding the Modal Decomposition-based Data Generation and Transform phases, the configuration is as follows. For the SVD algorithm, the number of modes provided has been set to $5$, considered adequate for the reconstruction according to the contribution of the modes in the frames. For the HODMD algorithm, the number of snapshots used on each sequence $K$ has been set to the total number of frames of the corresponding sequence. In this way, the maximum number of cardiac cycles possible can be captured and the cardiac and respiratory frequencies can be more accurately obtained. The time interval between snapshots $\Delta t = 4~ms$ and the tolerances $\epsilon_{SVD} = \epsilon_{DMD} = 5 \mathrm{e}{-4}$ are based on~\cite{Groun2022higher}. This is, $\Delta t$ is based on the used ultrasound scanner. Regarding the tolerances for the dimensionality reduction and amplitude truncation steps, respectively, these are larger than the noise level, and lead to a reasonable number of frequencies associated to the studied heart states. The index $d$ has been configured for each sequence in the way that fair reconstructions of the snapshots are obtained. In particular, for the diabetic cardiomyopathy, and healthy, depending on the sequence, $d = \left \lfloor K/3 \right \rfloor$ or $d = \left \lfloor K/5 \right \rfloor$ works better. Regarding the myocardial infarction and TAC hypertrophy sequences, $d = 50$ is considered adequate.

Concerning the Deep Neural Network-based Pathology Prediction phase, the input images have been resized to $256 \times 256$ and the patch size has been set to $32$. 

Finally, regarding the Fusion of Pathology Predictions phase, maximum and average scores have been used to calculate the global prediction score per class. This is, to calculate a global prediction score per class in a test sequence, the maximum of the predictions associated to a class is calculated to obtain the global prediction score of that class, or instead the average of the predictions.

\item Performance: Table \ref{tab:table_results_system_avg} and Table \ref{tab:table_results_system_max} reveal the impact of the training cases presented in Table \ref{tab:training_cases} using the different types of data for testing on the performance of the proposed system with and without the fusion of predictions via the global prediction score using the average and the maximum, respectively. The results show a general improvement in the training cases with respect to the case 1 (with only original images), considering that, in general, more training samples are available, as expected. However, the variation of the performance is also affected by the specific types of training data: for example, although having a similar number of training samples, the case 12 leads to much better performances than those with the case 9. The former uses SVD, HODMD modes and the SVD reconstructions of the HODMD modes, but the latter uses the original images instead. Also note that there is a consistent improvement in the training cases with HODMD modes, and also with their corresponding SVD reconstructions, i.e., cases 7, 10, 11, and 12. Therefore, these two facts show the importance of using HODMD modes for training. This is because the HODMD modes describe the dynamics of the data and contain the temporal information, leveraged by the neural network, which could be relevant to distinguish the different heart states. On the other hand, among the types of test data, a general improvement is produced with the reconstructed images. This is because these have less noise than the original data and, therefore, contain more discriminative features associated to the different heart states. In addition, there are more reconstructed images than modes in both training and testing, which implies that the neural network can better learn from image reconstructions, and also that there are more test data in each sequence in order to more reliably determine its heart state. So, this shows the effectiveness of the SVD and HODMD algorithms for data augmentation and extraction of more discriminative features associated to the different heart states.

\end{enumerate}

\begin{table*}[!ht]
\centering
\caption{Comparison results with different types of data for test and training cases using the proposed framework and the average score value from the predictions per class. Refer to Table~\ref{tab:training_cases} for the specifications of each training case.}
\label{tab:table_results_system_avg}
\resizebox{\textwidth}{!}{%
\begin{tabular}{@{}cccccccccccccc@{}}
\toprule
                       &                & \multicolumn{12}{c}{\textbf{Per-image accuracy (w/o)}}                                                               \\ \midrule
                       &               & Case 1 & Case 2 & Case 3 & Case 4 & Case 5 & Case 6 & Case 7 & Case 8 & Case 9 & Case 10 & Case 11 & Case 12         \\ \cmidrule{3-14}
                       & Original       & 0.49   & 0.56   & 0.54   & 0.56   & 0.57   & 0.48   & 0.55   & 0.59   & 0.52   & 0.54    & 0.55    & 0.62            \\
\multirow{2}{*}{SVD}   & Reconstruction & 0.52   & 0.57   & 0.53   & 0.59   & 0.56   & 0.50   & 0.55   & 0.58   & 0.52   & 0.54    & 0.55    & 0.63            \\
                       & Modes          & 0.24   & 0.17   & 0.38   & 0.18   & 0.35   & 0.23   & 0.38   & 0.22   & 0.14   & 0.42    & 0.40    & 0.47            \\
\multirow{2}{*}{HODMD} & Reconstruction & 0.53   & 0.56   & 0.52   & 0.58   & 0.54   & 0.50   & 0.57   & 0.58   & 0.51   & 0.56    & 0.56    & $\mathbf{0.64}$ \\
                       & Modes (abs)    & 0.39   & 0.38   & 0.45   & 0.41   & 0.48   & 0.40   & 0.59   & 0.40   & 0.37   & 0.53    & 0.57    & 0.61            \\
                       \midrule
                       &                & \multicolumn{12}{c}{\textbf{Per-image accuracy (w/): Average}}                                                       \\
                       \midrule
                       & Original       & 0.52   & 0.60   & 0.57   & 0.60   & 0.64   & 0.51   & 0.57   & 0.64   & 0.60   & 0.57    & 0.64    & 0.64            \\
\multirow{2}{*}{SVD}   & Reconstruction & 0.56   & 0.57   & 0.57   & 0.57   & 0.60   & 0.51   & 0.57   & 0.64   & 0.57   & 0.57    & 0.60    & 0.68            \\
                       & Modes          & 0.27   & 0.17   & 0.53   & 0.13   & 0.53   & 0.23   & 0.60   & 0.17   & 0.10   & 0.47    & 0.50    & 0.57            \\
\multirow{2}{*}{HODMD} & Reconstruction & 0.60   & 0.56   & 0.49   & 0.56   & 0.56   & 0.49   & 0.60   & 0.64   & 0.52   & 0.60    & 0.56    & $\mathbf{0.68}$ \\
                       & Modes (abs)    & 0.48   & 0.50   & 0.50   & 0.50   & 0.52   & 0.43   & 0.64   & 0.54   & 0.42   & 0.61    & 0.60    & 0.64            \\
                       \midrule
                       &                & \multicolumn{12}{c}{\textbf{Per-sequence accuracy: Average}}                                                         \\
                       \midrule
                       & Original       & 0.60   & 0.57   & 0.53   & 0.57   & 0.73   & 0.57   & 0.67   & 0.73   & 0.70   & 0.63    & 0.73    & 0.73            \\
\multirow{2}{*}{SVD}   & Reconstruction & 0.63   & 0.67   & 0.67   & 0.67   & 0.70   & 0.57   & 0.67   & 0.73   & 0.67   & 0.63    & 0.70    & 0.77            \\
                       & Modes          & 0.27   & 0.17   & 0.53   & 0.13   & 0.53   & 0.23   & 0.60   & 0.17   & 0.10   & 0.47    & 0.50    & 0.57            \\
\multirow{2}{*}{HODMD} & Reconstruction & 0.69   & 0.65   & 0.58   & 0.65   & 0.65   & 0.58   & 0.69   & 0.73   & 0.62   & 0.65    & 0.65    & $\mathbf{0.77}$ \\
                       & Modes (abs)    & 0.50   & 0.54   & 0.54   & 0.54   & 0.58   & 0.46   & 0.69   & 0.58   & 0.46   & 0.65    & 0.65    & 0.69            \\
  \cmidrule{2-14} &
  \# Samples (train) &
  \num[group-separator={~}]{9326} &
  \num[group-separator={~}]{9326} &
  \num[group-separator={~}]{19047} &
  \num[group-separator={~}]{18652} &
  \num[group-separator={~}]{9721} &
  \num[group-separator={~}]{9223} &
  \num[group-separator={~}]{13588} &
  \num[group-separator={~}]{18549} &
  \num[group-separator={~}]{27875} &
  \num[group-separator={~}]{17953} &
  \num[group-separator={~}]{23309} &
  \num[group-separator={~}]{27674} \\ \bottomrule
\end{tabular}%
}
\end{table*}

\begin{table*}[!ht]
\centering
\caption{Comparison results with different types of data for test and training cases using the proposed framework and the maximum score value from the predictions per class. Refer to Table~\ref{tab:training_cases} for the specifications of each training case.}
\label{tab:table_results_system_max}
\resizebox{\textwidth}{!}{%
\begin{tabular}{@{}cccccccccccccc@{}}
\toprule
                        &                & \multicolumn{12}{c}{\textbf{Per-image accuracy (w/o)}}                                                               \\ \midrule
                       &                & Case 1 & Case 2 & Case 3 & Case 4 & Case 5 & Case 6 & Case 7 & Case 8 & Case 9 & Case 10 & Case 11 & Case 12         \\ \cmidrule{3-14}
                       & Original       & 0.49   & 0.56   & 0.54   & 0.56   & 0.57   & 0.48   & 0.55   & 0.59   & 0.52   & 0.54    & 0.55    & 0.62            \\
\multirow{2}{*}{SVD}   & Reconstruction & 0.52   & 0.57   & 0.53   & 0.59   & 0.56   & 0.50   & 0.55   & 0.58   & 0.52   & 0.54    & 0.55    & 0.63            \\
                       & Modes          & 0.24   & 0.17   & 0.38   & 0.18   & 0.35   & 0.23   & 0.38   & 0.22   & 0.14   & 0.42    & 0.40    & 0.47            \\
\multirow{2}{*}{HODMD} & Reconstruction & 0.53   & 0.56   & 0.52   & 0.58   & 0.54   & 0.50   & 0.57   & 0.58   & 0.51   & 0.56    & 0.56    & $\mathbf{0.64}$ \\
                       & Modes (abs)    & 0.39   & 0.38   & 0.45   & 0.41   & 0.48   & 0.40   & 0.59   & 0.40   & 0.37   & 0.53    & 0.57    & 0.61            \\
                       \midrule
                       &                & \multicolumn{12}{c}{\textbf{Per-image accuracy (w/): Maximum}}                                                       \\
                       \midrule
                       & Original       & 0.48   & 0.64   & 0.57   & 0.57   & 0.63   & 0.51   & 0.57   & 0.64   & 0.57   & 0.64    & 0.55    & 0.64            \\
\multirow{2}{*}{SVD}   & Reconstruction & 0.52   & 0.60   & 0.53   & 0.57   & 0.59   & 0.51   & 0.57   & 0.64   & 0.53   & 0.61    & 0.60    & 0.64            \\
                       & Modes          & 0.20   & 0.13   & 0.57   & 0.17   & 0.67   & 0.20   & 0.53   & 0.20   & 0.13   & 0.50    & 0.60    & 0.53            \\
\multirow{2}{*}{HODMD} & Reconstruction & 0.56  & 0.60   & 0.49   & 0.56   & 0.56   & 0.52   & 0.56   & 0.64   & 0.52   & 0.60    & 0.56    & $\mathbf{0.68}$ \\
                       & Modes (abs)    & 0.22   & 0.36   & 0.51   & 0.36   & 0.63   & 0.45   & 0.57   & 0.49   & 0.36   & 0.61    & 0.60    & 0.60            \\
                       \midrule
                       &                & \multicolumn{12}{c}{\textbf{Per-sequence accuracy: Maximum}}                                                         \\
                       \midrule
                       & Original       & 0.57   & 0.60   & 0.53   & 0.53   & 0.70   & 0.57   & 0.67   & 0.73   & 0.67   & 0.70    & 0.63    & 0.73            \\
\multirow{2}{*}{SVD}   & Reconstruction & 0.60   & 0.70   & 0.63   & 0.67   & 0.67   & 0.57   & 0.67   & 0.73   & 0.63   & 0.67    & 0.70    & 0.73            \\
                       & Modes          & 0.20   & 0.13   & 0.57   & 0.17   & 0.67   & 0.20   & 0.53   & 0.20   & 0.13   & 0.50    & 0.60    & 0.53            \\
\multirow{2}{*}{HODMD} & Reconstruction & 0.65   & 0.69   & 0.58   & 0.65   & 0.65   & 0.62   & 0.65   & 0.73   & 0.62   & 0.65    & 0.65    & $\mathbf{0.77}$ \\
                       & Modes (abs)    & 0.23   & 0.38   & 0.54   & 0.38   & 0.69   & 0.50   & 0.62   & 0.54   & 0.38   & 0.65    & 0.65    & 0.65            \\
  \cmidrule{2-14} &
  \# Samples (train) &
  \num[group-separator={~}]{9326} &
  \num[group-separator={~}]{9326} &
  \num[group-separator={~}]{19047} &
  \num[group-separator={~}]{18652} &
  \num[group-separator={~}]{9721} &
  \num[group-separator={~}]{9223} &
  \num[group-separator={~}]{13588} &
  \num[group-separator={~}]{18549} &
  \num[group-separator={~}]{27875} &
  \num[group-separator={~}]{17953} &
  \num[group-separator={~}]{23309} &
  \num[group-separator={~}]{27674} \\ \bottomrule
\end{tabular}%
}
\end{table*}

Table \ref{tab:table_results_per_class} presents the recognition results for each heart state using the best configuration inferred from Table \ref{tab:table_results_system_avg} and Table \ref{tab:table_results_system_max}: training case 12 and HODMD reconstructions as test data with and without fusion of predictions using the average and or the maximum. The performance is balanced among the four considered heart states, like the number of test samples in terms of both images and sequences. A 77\% average per-sequence accuracy is reached, which is high enough considering the low number of training samples (images) from each class, much lower than the samples required by typical deep learning algorithms (hundreds of thousands or even millions of samples). Therefore, with a much higher number of samples, the different heart states could be better recognized and the associated features better learned. 

\begin{table*}[!ht]
\centering
\caption{Cardiac pathology recognition performance using F1-Score ($\mathit{F}$) and accuracy obtained with the best configuration of the proposed framework. Refer to Table~\ref{tab:training_cases} for the specifications of each training case.}
\label{tab:table_results_per_class}
\resizebox{\textwidth}{!}{%
\begin{tabular}{@{}cccccccc@{}}
\toprule
            &                    & \multicolumn{2}{c}{$\mathit{F}$ (Average)} & \multicolumn{2}{c}{$\mathit{F}$ (Maximum)} & \multicolumn{2}{c}{Test Data} \\ \midrule
Heart state & $\mathit{F}$ (w/o) & (w/)                 & Per-sequence        & (w/)             & Per-sequence            & \# Images    & \# Sequences   \\ \midrule
Diabetic Cardiomyopathy & 0.64 & 0.63 & 0.71 & 0.63 & 0.71 & 715 & 7  \\
Healthy                 & 0.64 & 0.72 & 0.86 & 0.72 & 0.86 & 837 & 7  \\
Myocardial Infarction   & 0.66 & 0.70 & 0.77 & 0.70 & 0.77 & 900 & 6  \\
TAC Hypertrophy        & 0.56 & 0.64 & 0.73 & 0.64 & 0.73 & 720 & 6  \\ \midrule
Accuracy    & $\mathbf{0.64}$    & $\mathbf{0.68}$      & $\mathbf{0.77}$     & 0.68             & 0.77                    & 3276         & 26             \\ \bottomrule
\end{tabular}%
}
\end{table*}

\subsection{Comparison With Alternative Algorithms}
\label{sec:analysis_algorithms}

In this second part of the experiments, the performance of the proposed system has been compared against other alternative approaches, but still using the same framework, where the ViT has been replaced by other deep neural networks. In particular, the comparison includes a CNN trained from scratch, and the pretrained models Inception-v3, ResNet50-v2, and VGG16. First, the configuration of the proposed system is introduced. Next, the configuration and the necessary adaptations of the alternative algorithms are described. Finally, an overall performance comparison between the proposed ViT and other deep neural networks is presented, using the best training cases (1, 7, 10, 11, 12) and the main types of test data (Original and HODMD-based reconstructions).

\begin{enumerate}

\item Configuration and adaptation of the alternative algorithms: Regarding the configuration of the SVD and HODMD algorithms, the ViT, and of the prediction, it is the same as in Subsection \ref{sec:analysis_data}. However, in the Fusion of Pathology Predictions phase, only the average scores are used to calculate the global prediction score per class in a test sequence.   

Before presenting the parameters selected for the CNN, Inception-v3, ResNet50-v2, and VGG16, some considerations must be taken into account. The input of these deep neural networks has three channels, but the cardiac pathology database used in this work comprises one-channel data. In addition, their architectures were designed to classify into the classes from the ImageNet challenge (\cite{russakovsky2015imagenet}). Therefore, adaptations were required to enable them to not only work with one-channel data, but also to classify into the heart states from the cardiac pathology database. To obtain three-channel data, the input has been replicated three times and concatenated in the channel dimension before being introduced into the deep neural networks. For classification, the final layer of these models is replaced by a series of dense layers as follows. In the Inception-v3 and VGG16 models, features are flattened, and two dense layers are introduced: the first one has $128$ elements and the second one the number of heart states. In the case of ResNet50-v2, a single dense layer, whose amount of elements is the number of heart states, is used. In addition, global average pooling is applied to the output of the last convolutional layer. The activation function after every dense layer is a rectifier linear unit (ReLU), except from the final one, which is a Softmax. For the initialization of Inception-v3, ResNet50-v2, and VGG16, pretrained weights from the ImageNet challenge (\cite{russakovsky2015imagenet}) have been used, instead of random ones. These pretrained weights give a good initialization, and therefore improve the training convergence and the quality of the models obtained after fine-tuning using the cardiac pathology database. This means that these algorithms use an auxiliary dataset in addition to the cardiac pathology database. As the proposed ViT and the CNN do not use that auxiliary dataset, it supposes an advantage for the algorithms.

The CNN is composed of three convolutional layers, each one followed by a max pooling layer. The activation function after every convolutional layer is a ReLU, and the number of filters have been set to $16$, $32$, and $64$, respectively. The sizes are $3 \times 3$ and $2 \times 2$ for the convolutional and max pooling layers, respectively. After that, features are flattened, and two dense layers are introduced: the first one has $128$ elements and the second one the number of heart states. The activation function of the dense layers are a ReLU and a Softmax, respectively.   

Similarly to the proposed ViT, the input images have been resized to $256 \times 256$. This image resizing is also mandatory due to the memory limitations of the GPU devices for training.

To reduce randomness in the results because of the training convergence of the algorithms in each experiment, each algorithm is trained four times using the configuration explained above. Then, results have been averaged, and the standard deviation of the results has been calculated.

\begin{table*}[!ht]
\centering
\caption{Comparison of the recognition performance among different algorithms with the main types of data and training cases using the average score value from the predictions per class. Refer to Table~\ref{tab:training_cases} for the specifications of each training case.}
\label{tab:table_results_comp}
\resizebox{\textwidth}{!}{%
\begin{tabular}{@{}ccccccc@{}}
\toprule
                              &                      & \multicolumn{5}{c}{\textbf{Per-image accuracy (w/o)}}                                                     \\ \midrule
Algorithm                     & Test Data            & Case 1          & Case 7                   & Case 10         & Case 11                  & Case 12         \\ \midrule 
\multirow{2}{*}{CNN}          & Original             & 0.40 $\pm$ 0.11 & 0.35 $\pm$ 0.17          & 0.44 $\pm$ 0.10 & 0.29 $\pm$ 0.05          & 0.45 $\pm$ 0.12 \\
                              & HODMD Reconstruction & 0.36 $\pm$ 0.11 & 0.37 $\pm$ 0.20          & 0.50 $\pm$ 0.16 & 0.34 $\pm$ 0.14          & 0.51 $\pm$ 0.19 \\
\multirow{2}{*}{Inception-v3} & Original             & 0.49 $\pm$ 0.01 & 0.49 $\pm$ 0.03          & 0.45 $\pm$ 0.06 & 0.42 $\pm$ 0.09          & 0.45 $\pm$ 0.08 \\
                              & HODMD Reconstruction & 0.44 $\pm$ 0.02 & 0.51 $\pm$ 0.04          & 0.45 $\pm$ 0.09 & 0.42 $\pm$ 0.12          & 0.46 $\pm$ 0.09 \\
\multirow{2}{*}{ResNet50-v2}  & Original             & 0.47 $\pm$ 0.01 & 0.46 $\pm$ 0.07          & 0.48 $\pm$ 0.05 & 0.50 $\pm$ 0.05          & 0.46 $\pm$ 0.06 \\
                              & HODMD Reconstruction & 0.45 $\pm$ 0.03 & 0.54 $\pm$ 0.03          & 0.56 $\pm$ 0.03 & 0.58 $\pm$ 0.01          & 0.54 $\pm$ 0.02 \\
\multirow{2}{*}{VGG16}        & Original             & 0.45 $\pm$ 0.10 & 0.39 $\pm$ 0.03          & 0.30 $\pm$ 0.09 & 0.38 $\pm$ 0.07          & 0.45 $\pm$ 0.05 \\
                              & HODMD Reconstruction & 0.33 $\pm$ 0.06 & 0.47 $\pm$ 0.05          & 0.32 $\pm$ 0.12 & 0.44 $\pm$ 0.13          & 0.54 $\pm$ 0.06 \\
\multirow{2}{*}{ViT}          & Original             & 0.52 $\pm$ 0.02 & 0.56 $\pm$ 0.03          & 0.56 $\pm$ 0.03 & 0.57 $\pm$ 0.03          & 0.56 $\pm$ 0.04 \\
                              & HODMD Reconstruction & 0.54 $\pm$ 0.02 & $\mathbf{0.58 \pm 0.04}$ & 0.57 $\pm$ 0.03 & 0.58 $\pm$ 0.03          & 0.57 $\pm$ 0.05 \\ \midrule
                              &                      & \multicolumn{5}{c}{\textbf{Per-image accuracy (w/): Average}}                                             \\ \midrule
\multirow{2}{*}{CNN}          & Original             & 0.38 $\pm$ 0.15 & 0.36 $\pm$ 0.24          & 0.48 $\pm$ 0.11 & 0.28 $\pm$ 0.08          & 0.49 $\pm$ 0.16 \\
                              & HODMD Reconstruction & 0.39 $\pm$ 0.12 & 0.41 $\pm$ 0.16          & 0.52 $\pm$ 0.21 & 0.30 $\pm$ 0.11          & 0.55 $\pm$ 0.21 \\
\multirow{2}{*}{Inception-v3} & Original             & 0.52 $\pm$ 0.05 & 0.53 $\pm$ 0.05          & 0.54 $\pm$ 0.10 & 0.46 $\pm$ 0.11          & 0.59 $\pm$ 0.09 \\
                              & HODMD Reconstruction & 0.47 $\pm$ 0.06 & 0.58 $\pm$ 0.08          & 0.46 $\pm$ 0.12 & 0.46 $\pm$ 0.09          & 0.52 $\pm$ 0.09 \\
\multirow{2}{*}{ResNet50-v2}  & Original             & 0.48 $\pm$ 0.03 & 0.53 $\pm$ 0.14          & 0.53 $\pm$ 0.10 & 0.57 $\pm$ 0.03          & 0.50 $\pm$ 0.12 \\
                              & HODMD Reconstruction & 0.47 $\pm$ 0.06 & 0.63 $\pm$ 0.06          & 0.63 $\pm$ 0.05 & 0.65 $\pm$ 0.03          & 0.63 $\pm$ 0.05 \\
\multirow{2}{*}{VGG16}        & Original             & 0.51 $\pm$ 0.12 & 0.39 $\pm$ 0.08          & 0.30 $\pm$ 0.08 & 0.40 $\pm$ 0.09          & 0.46 $\pm$ 0.07 \\
                              & HODMD Reconstruction & 0.33 $\pm$ 0.07 & 0.47 $\pm$ 0.08          & 0.33 $\pm$ 0.11 & 0.49 $\pm$ 0.17          & 0.57 $\pm$ 0.06 \\
\multirow{2}{*}{ViT}          & Original             & 0.55 $\pm$ 0.03 & 0.58 $\pm$ 0.02          & 0.57 $\pm$ 0.06 & 0.63 $\pm$ 0.02          & 0.61 $\pm$ 0.04 \\
                              & HODMD Reconstruction & 0.57 $\pm$ 0.03 & 0.61 $\pm$ 0.02          & 0.59 $\pm$ 0.07 & 0.61 $\pm$ 0.04          & 0.61 $\pm$ 0.06 \\ \midrule
                              &                      & \multicolumn{5}{c}{\textbf{Per-sequence accuracy (w/): Average}}                                          \\ \midrule
\multirow{2}{*}{CNN}          & Original             & 0.49 $\pm$ 0.14 & 0.38 $\pm$ 0.30          & 0.57 $\pm$ 0.14 & 0.31 $\pm$ 0.07          & 0.57 $\pm$ 0.20 \\
                              & HODMD Reconstruction & 0.42 $\pm$ 0.13 & 0.38 $\pm$ 0.25          & 0.59 $\pm$ 0.24 & 0.31 $\pm$ 0.11          & 0.63 $\pm$ 0.22 \\
\multirow{2}{*}{Inception-v3} & Original             & 0.59 $\pm$ 0.07 & 0.61 $\pm$ 0.07          & 0.59 $\pm$ 0.10 & 0.55 $\pm$ 0.12          & 0.68 $\pm$ 0.10 \\
                              & HODMD Reconstruction & 0.50 $\pm$ 0.07 & 0.62 $\pm$ 0.11          & 0.51 $\pm$ 0.15 & 0.48 $\pm$ 0.12          & 0.57 $\pm$ 0.12 \\
\multirow{2}{*}{ResNet50-v2}  & Original             & 0.56 $\pm$ 0.03 & 0.60 $\pm$ 0.14          & 0.63 $\pm$ 0.11 & 0.65 $\pm$ 0.06          & 0.58 $\pm$ 0.12 \\
                              & HODMD Reconstruction & 0.50 $\pm$ 0.03 & 0.67 $\pm$ 0.05          & 0.67 $\pm$ 0.05 & 0.66 $\pm$ 0.02          & 0.69 $\pm$ 0.07 \\
\multirow{2}{*}{VGG16}        & Original             & 0.54 $\pm$ 0.15 & 0.42 $\pm$ 0.08          & 0.34 $\pm$ 0.12 & 0.43 $\pm$ 0.09          & 0.50 $\pm$ 0.09 \\
                              & HODMD Reconstruction & 0.30 $\pm$ 0.09 & 0.51 $\pm$ 0.10          & 0.38 $\pm$ 0.12 & 0.51 $\pm$ 0.21          & 0.62 $\pm$ 0.07 \\
\multirow{2}{*}{ViT}          & Original             & 0.68 $\pm$ 0.05 & 0.68 $\pm$ 0.02          & 0.65 $\pm$ 0.07 & $\mathbf{0.72 \pm 0.02}$ & 0.70 $\pm$ 0.05 \\
                              & HODMD Reconstruction & 0.65 $\pm$ 0.03 & 0.70 $\pm$ 0.02          & 0.66 $\pm$ 0.08 & 0.70 $\pm$ 0.04          & 0.70 $\pm$ 0.07 \\  \cmidrule{2-7}
 &
  \# Samples (train) &
  $\num[group-separator={~}]{9326}$ &
  $\num[group-separator={~}]{13588}$ &
  $\num[group-separator={~}]{17953}$ &
  $\num[group-separator={~}]{23309}$ &
  $\num[group-separator={~}]{27674}$ \\ \bottomrule
\end{tabular}%
}
\end{table*}

\item Performance: Table \ref{tab:table_results_comp} presents a comparison of the cardiac pathology recognition performance, using the medical database and the proposed system with either the ViT, the CNN, Inception-v3, ResNet50-v2, or VGG16. Only the training cases 1, 7, 10, 11, and 12 have been considered, as being the most representative of the contribution of the SVD and HODMD algorithms with respect to the use of only original images for training. Regarding the types of test data, original images and HODMD-based reconstructions have been compared, as mainly representing the influence of the SVD and HODMD algorithms as feature extractors.

As can be observed in the table, the two best results in terms of recognition performance have been obtained using the proposed ViT and the ResNet50-v2, according to the mean and standard deviations of the accuracies. Moreover, these even outperform the CNN without pretraining, showing the efficacy of using more complex models, with techniques to better train them (e.g., residual networks, or the SPT and LSA modules in the proposed ViT). In addition, there is a general improvement in the training cases with respect to the case 1 and in the HODMD-based reconstructions for test data. Therefore, the alternative algorithms also benefit from the use of the SVD and HODMD algorithms for both data augmentation and feature extraction, also incorporating temporal information thanks to the HODMD modes. Note that, in the training case 1, a better performance is generally achieved when using original test images. This can be explained by the fact that the models learn to recognize different heart states in noisy images instead of in filtered ones, obtaining more confident predictions in the original images.      

Regarding the per-image accuracy (w/o), the proposed ViT achieves the best results. Nonetheless, ResNet50-v2 performs very closely to it. This is mainly because ResNet50-v2 has been pretrained with the external ImageNet database, which contributes to compensate for the fact that the cardiac pathology database has relatively a lower number of samples in comparison with other databases used by deep learning techniques. In this sense, a larger cardiac pathology database could significantly improve the proposed framework with the alternative models, trained from scratch or not. In particular, the training case 7 has led to the best results in the ViT, showing the improvement given by the HODMD-based data, but with practically no impact when incorporating SVD-based data, as possibly barely giving new training data (case 11). On the contrary, the per-sequence accuracy is the best one with more difference in the proposed ViT, and specifically in the training case 11, showing that the images which have been correctly classified lead to more correctly classified sequences. In addition, a higher performance is obtained in the case of using original images in test. Once more, this is due to that the distribution and confidences of the images correctly classified among the test sequences are more favourable than in the case of HODMD test data, obtaining more correctly classified test sequences.  

Computational cost of the different phases of the prediction process is summarized in Table \ref{tab:comp_cost}. It is important to note that the SVD algorithms and the HODMD algorithms are applied on sequences rather than on single images, but the time measures represented in the table are averages per image. As can be seen, the HOSVD dimensionality reduction process is the hardest part by far (\cite{Bell2023HODMD}). Although all the compared deep learning algorithms have very low prediction times, the proposed ViT is the fastest one, which is desirable considering the lower number of parameters and the better recognition performance. This result also encourages the deployment of the proposed ViT using less resources. In every case, all time values are inside of real-time requirements, which is appealing for several medical applications.

\begin{table}[!ht]
\centering
\caption{Computational cost of different phases of the cardiac pathology recognition system using the algorithms compared.}
\label{tab:comp_cost}
\resizebox{10cm}{!}{%
\begin{tabular}{@{}cccccc@{}}
\toprule
Model & \# Parameters                        & $\bar{t}_{SVD}$      & $\bar{t}_{HOSVD}$    & $\bar{t}_{HODMD}$      & $\bar{t}_{pred}$ \\ \midrule
CNN   & $\num[group-separator={~}]{7396872}$ & \multirow{5}{*}{5.1} & \multirow{5}{*}{591} & \multirow{5}{*}{0.648} & 2                \\
Inception-v3 & $\num[group-separator={~}]{31240744}$         &  &  &  & 3              \\
ResNet50-v2  & $\num[group-separator={~}]{23575048}$         &  &  &  & 3.2            \\
VGG16        & $\num[group-separator={~}]{18909768}$         &  &  &  & 4.1            \\
ViT          & $\mathbf{\num[group-separator={~}]{5730512}}$ &  &  &  & $\mathbf{1.5}$ \\ 
\bottomrule
\end{tabular}%
}
\end{table}

For short, the best overall performance is achieved with the proposed ViT in terms of both recognition per-image and per-sequence accuracy and operation time. Lastly, recognition accuracy is expected to improve with larger databases, since it is trained from scratch, unlike Inception-v3, ResNet50-v2, and VGG16.

\end{enumerate}

\section{Conclusions}
\label{sec:conclusions}

Heart diseases are the main cause of human mortality in the world. This makes their recognition a task of great importance. In this work, a real-time cardiac pathology recognition system for echocardiography video sequences has been presented to face this challenging and demanding task. The two major contributions of this paper are the creation of a large annotated medical database from echocardiography sequences with heterogeneous acquisition features, and a deep neural network, based on a ViT, designed for an effective training from scratch, even with small datasets. During the creation of the database, the HODMD algorithm has been employed for the first time (to the best of the authors' knowledge) as a feature extractor and a data augmentation technique in the medical field, which has demonstrated to improve the cardiac pathology recognition performance. The results obtained have proved that the proposed system outperforms CNNs, even with pretraining. We anticipate that, if longer datasets were available, overfitting and robustness would be better.

\section*{Declaration of competing interest}
None Declared.

\section*{Acknowledgements}
This work was supported by Grants TED2021-129774B-C21, PID2021-124629OB-I00, TED2021-129774B-C22, and PLEC2022-009235, funded by the Ministry of Science and Innovation (MCIN/AEI/10.13039/501100011033), by the European Union's NextGenerationEU/PRTR ("Plan de Recuperación, Transformación y Resiliencia de España"), and by FEDER: the first one to A.B-N and N.G and the other three to E.L-P, by Grant PEJ-2019-TL/BMD-12831 from Comunidad de Madrid to E.L-P and to M.V-O, and by a Juan de la Cierva Incorporación Grant (IJCI-2016-27698) to M.V-O. The CNIC is supported by the Instituto de Salud Carlos III (ISCIII), the Ministerio de Ciencia e Innovación (MCIN), and the Pro CNIC Foundation, and is a Severo Ochoa Center of Excellence (Grant CEX2020-001041-S funded by MICIN/AEI/10.13039/501100011033).

\bibliographystyle{elsarticle-harv} 
\bibliography{bibliography}





\end{document}